\documentclass[prd,aps,nofootinbib,superscriptaddress]{revtex4}
\usepackage{epsfig}
\usepackage{amsmath, slashed}
\usepackage{color}
\usepackage{amsfonts}
\usepackage{enumitem}

\usepackage[mathscr]{eucal}

\newcommand{\ucal}{\mathcal}
\newcommand{\uvec}{\boldsymbol}
\newcommand{\ud}{\mathrm{d}}

\newcommand{\beq}{\begin{equation}}
\newcommand{\eeq}{\end{equation}}
\newcommand{\beqyn}{\begin{eqnarray*}}
\newcommand{\eeqyn}{\end{eqnarray*}}

\newcommand{\la}{\langle}
\newcommand{\ra}{\rangle}
\newcommand{\barpsi}{\overline{\psi}}
\newcommand{\baru}{\overline{u}}
\newcommand{\LRD}{\overset{\leftrightarrow}{D}\!\!\!\!\!\phantom{D}}

\newlength\savedwidth

\newcommand\whline{\noalign{\global\savedwidth\arrayrulewidth
\global\arrayrulewidth 1pt}%
\hline
\noalign{\global\arrayrulewidth\savedwidth}}

\allowdisplaybreaks
\begin{document}

\title{Relativistic spin sum rules and the role of the pivot}
 
\author{C\'edric Lorc\'e}
\affiliation{CPHT, CNRS, Ecole Polytechnique, Institut Polytechnique de Paris, Route de Saclay, 91128 Palaiseau, France}



\begin{abstract}
Spin sum rules depend on the choice of a pivot, i.e. the point about which the angular momentun is defined, usually identified with the center of the nucleon. The latter is however not unique in a relativistic theory and has led to apparently contradictory results in the literature. Using the recently developed phase-space approach, we compute for the first time the contribution associated with the motion of the center of the nucleon, and we derive a general spin sum rule which reduces to established results after appropriate choices for the pivot and the spin component.
\end{abstract}

\maketitle

\section{Introduction}

After more than three decades of intense study on both theoretical and experimental sides, the spin structure of the nucleon remains one of the most vivid topics in hadronic physics. Reviews about key developments and achievements in this field can be found in Refs.~\cite{Filippone:2001ux,Bass:2004xa,Aidala:2012mv,Leader:2013jra,Wakamatsu:2014zza,Liu:2015xha,Deur:2018roz,Ji:2020ena}. A fundamental question is the decomposition of the nucleon spin into its constituents' intrinsic and orbital angular momentum contributions. While longitudinal spin (or helicity) sum rules derived long ago~\cite{Jaffe:1989jz,Ji:1996ek} are well-established, various transverse spin sum rules have been discussed over the years and led to a rather confused situation~\cite{Burkardt:2002hr,Bakker:2004ib,Burkardt:2005hp,Leader:2011cr,Ji:2012sj,Ji:2012vj,Leader:2012md,Harindranath:2012wn,Leader:2012ar,Hatta:2012jm,Harindranath:2013goa,Ji:2013tva}. Deriving a transverse spin sum rule is a delicate and subtle problem for the simple reason that transverse rotations (unlike longitudinal rotations) do not commute with longitudinal boosts. In the review~\cite{Leader:2013jra}, a whole section has been devoted to discuss in detail these spin sum rules and to clarify their status. 

In a recent paper~\cite{Ji:2020hii} a number of results about the transverse angular momentum (AM) in a transversely polarized nucleon are derived, and seem to contradict those obtained earlier in Refs.~\cite{Bakker:2004ib,Leader:2011cr}. The authors argue that the old results are ``incorrect'' because they contain, in a moving frame, contributions from the center of mass motion. To get rid of these contributions, they observe that some terms in the expectation value of the transverse AM do not contribute to the Pauli-Luba\'nski pseudovector, and accordingly remove them by hand. This prescription appears however somewhat ad hoc and requires rigorous justification. In particular, the contribution from the center-of-mass motion is actually never calculated.

In this paper, we show that the old~\cite{Bakker:2004ib,Leader:2011cr} and the new~\cite{Ji:2020hii} results are in fact both correct. They just refer to different definitions for the ``center'' of the nucleon, and hence for what is meant by ``spin'' or ``internal AM''. The key role played by this center was recognized long ago~\cite{Bakker:2004ib,Burkardt:2002hr,Burkardt:2005hp,Lorce:2011kn} and motivated a recent discussion of the concept of relativistic center of mass in the context of hadronic physics~\cite{Lorce:2018zpf}, which clarified the leading-twist picture of AM in the light-front formalism addressed in Refs.~\cite{Hoodbhoy:1998yb,Burkardt:2002hr,Burkardt:2005hp,Ji:2012sj,Ji:2012vj}. Here we adopt the phase-space approach developed in Refs.~\cite{Lorce:2018zpf,Lorce:2017wkb,Lorce:2018egm} to identify and compute explicitly the contribution from the center-of-mass motion. In the case of the transverse AM, we show that adding this contribution is equivalent to the somewhat ad hoc prescription used in Ref.~\cite{Ji:2020hii}. We observe however that this prescription applies only to the AM operators and leads to incorrect results for the transverse boost operators.  

The rest of this paper is organized as follows. In Sec.~\ref{sec:spin} we review the notion of spin in Quantum Field Theory and the problem of defining the center of the nucleon. Then we explain in Sec.~\ref{sec:ME} how to compute rigorously the expectation values for operators of the form $\int\ud^3r\,r^iO(r)$ and apply the formalism to the Poincar\'e generators in Sec.~\ref{sec:AMSR}, followed by a derivation of the general spin sum rule and a comparison with former results found in the literature. Finally, we summarize our findings in Sec.~\ref{sec:conclusions}.

\section{What do we mean by ``spin''?}\label{sec:spin}

An AM operator is usually understood as a generator of rotations about some point, the pivot. In practice, once a Lorentz frame is fixed, the origin of the coordinate system is often chosen to coincide with the pivot. But the choice of the pivot is arbitrary. It is in general an abstract concept unless it is given some physical meaning. For example, in the simplest non-relativistic treatment of the hydrogen atom the pivot is placed at the nucleus, which is taken to be an infinitely heavy point charge, with the spinless electron moving in its Coulomb potential.  

Problems with understanding the magnetic moment in the Stern-Gerlach experiment led to the introduction of ``spin'' in Quantum Mechanics to represent some new intrinsic property of the electron, eventually recognized as  a new type of AM with no classical analog. The concept of spin was then extended to composite systems and identified with the internal AM\footnote{Generally the internal AM will contain a mixture of contributions from both orbital motion and intrinsic AM of the constituents.}, defined as 
\begin{equation} \label{S}
\uvec S=\uvec J-\uvec R \times \uvec P,
\end{equation}
where $\uvec J$ is the total AM about the origin, $\uvec R$ is the ``position'' of the system with respect to that origin, and $\uvec P$ is its total momentum. 

If one requires $\uvec S$ to generate internal rotations, i.e. rotations about the position of the system, then the external AM operator $\uvec L=\uvec R\times\uvec P$ must generate external rotations about the origin, i.e. it must satisfy the following commutation relations
\begin{equation}
    [R^i,L^j]=i\epsilon^{ijk}R^k,\qquad [P^i,L^j]=i\epsilon^{ijk}P^k,
\end{equation}
and hence $[L^i,L^j]=i\epsilon^{ijk}L^k$. The necessary and sufficient conditions for these to hold are that the operators $\uvec R$ and $\uvec P$ obey the standard canonical commutation relations
\begin{equation}\label{AMcond}
    [R^i,R^j]=0,\qquad [P^i,P^j]=0,\qquad [R^i,P^j]=i\delta^{ij}.
\end{equation}
They imply in particular
\begin{equation}
    [R^i,S^j]=0,\qquad [P^i,S^j]=0,\qquad [S^i,S^j]=i\epsilon^{ijk}S^k.
\end{equation}
In other words, the internal AM operator $\uvec S$ satisfies the usual $su(2)$ algebra and is independent of the position and the total momentum of the system, and therefore of the external AM operator $\uvec L$.
\newline

The question of interest is: what point defines the position of the system? In non-relativistic Quantum Mechanics, it is defined by the center-of-mass position operator
\beq \label{NRCM} 
\uvec R := \uvec{ R}_\text{NR}=\frac{1}{m}\sum_{n=1}^N m_n\uvec R_n,
\eeq
where $m=\sum_n m_n$ is the mass of a system made of $N$ particles with masses $m_n$. Since the particles' position and momentum operators satisfy by definition $[R^i_n,P^j_{n'}]=i\delta^{ij}\delta_{nn'}$ with $[R^i_n,R^j_{n'}]=0$ and $[P^i_n,P^j_{n'}]=0$, one sees that the non-relativistic spin operator $\uvec S_\text{NR}=\uvec J-\uvec R_\text{NR} \times \uvec P$ is the generator of internal rotations. If the system consists of a single structureless particle, then the non-relativistic spin operator simply coincides with the intrinsic AM operator.

In a relativistic theory, the situation is more complicated for two reasons:
\begin{itemize}
    \item There exist different expressions for the AM operators in terms of the fields associated with the particles. The two principal versions are the Canonical and the Belinfante (or symmetrized) forms, which differ by a spatial integral over fields evaluated at infinity. As explained in Section 2.6.3 of Ref.~\cite{Leader:2013jra}, it would be impossible to construct a consistent theory if one could not ignore such terms in the matrix elements of these AM operators taken between normalizable states. Hence we may choose which version to use, and in this paper, as is done in the papers we are commenting on, we shall only use the Belinfante versions. 
    \item In particle physics it is often adequate to describe the motion of a particle as being in a momentum eigenstate, i.e. a plane wave, but clearly in that case the density in coordinate space is constant so that the particle is totally delocalized.  So, to discuss internal AM  some form of localized wave packet is essential. In addition, even when dealing with a localized system, there still exist several possibilities for the definition of the position of the system, which are identical in the rest frame of the system, but generally differ in other frames. The main three definitions are given by the so-called relativistic centers of energy (or inertia), mass, and spin~\cite{Pryce:1948pf,Moller:1949,Fleming:1965,Choi:2014nea,Lorce:2018zpf}, which we briefly review in the following.
\end{itemize}

\subsection{The relativistic center of energy or inertia}

The relativistic center of energy or inertia is the field theoretical version of Eq.~\eqref{NRCM}, where the role of inertial mass is played by the energy. Its position and time component are defined, in any reference frame, by~\cite{Fokker:1929,Born:1935ap}
\beq \label{RE} 
R^\mu_E=\frac{1}{P^0}\int \ud^3r \, r^\mu \, T^{00} (r),
\eeq
where the operator $T^{\mu\nu}(r)$ is the energy momentum tensor (EMT) and $P^\mu=\int\ud^3r\,T^{0\mu}(r)$ is the total four-momentum operator. Acting on a four-momentum eigenstate with eigenvalue $p^\mu$, the operator $1/P^0$ simply means multiplication by $1/p^0$. Also, ambiguities in the ordering of Hermitian operators are resolved by considering implicitly symmetrized products $AB\equiv\frac{1}{2}(AB+BA)$~\cite{Born:1935ap,Pryce:1948pf}.

In a field theory, the AM generators $J^i=\frac{1}{2} \epsilon^{ijk}M^{jk}$ and the generators of boosts $K^i=M^{0i}$ are defined in terms of the generalized AM momentum tensor, which reads in Belinfante form~\cite{Belinfante:1940}
\begin{equation} \label{GAM}
M^{\mu\nu}=-M^{\nu\mu}=\int\ud^3r\left[r^\mu T^{0\nu}(r)-r^\nu T^{0\mu}(r)\right].
\end{equation}
It follows that the position operator for the relativistic center of energy is directly related to the generators of boosts as follows
\beq  \label{K} 
\uvec R_E=t\,\frac{\uvec P}{P^0}-\frac{\uvec K}{P^0}. 
\eeq
Clearly, $R^\mu_E$ does not transform as a Lorentz four-vector. Using the following commutation relations
\beq \label{ComR} 
[K^i,K^j]= -i\epsilon^{ijk}J^k, \qquad [P^0,K^i]= i P^i, \qquad [P^i,K^j]=i\delta^{ij}P^0, 
\eeq
and the fact that $\uvec J$ generates the rotations, one can check that  $\uvec R_E$ satisfies the canonical commutation relations $[R^i_E,P^j]=i\delta^{ij}$ and $[R^i_E,J^j]=i\epsilon^{ijk}R^k_E$, so that it qualifies to some extent as a position operator, albeit with mutually non-commuting components. This last property implies that $\uvec R_E\times\uvec P$ will not satisfy the $su(2)$ algebra and cannot be considered as a genuine orbital AM operator. Nonetheless, as is customary, we shall continue to refer to it as the AM of the center of energy about the origin.

The nearest one can get to a spin operator transforming as a four-vector is the Pauli-Luba\'nski pseudovector~\cite{Lubanski:1942,Lubanski:1942bis}
\beq \label{PL} 
W^\mu= \frac{1}{2}\,\epsilon^{\mu \alpha \beta \lambda} M_{\alpha \beta} P_{\lambda},
\eeq
which satisfies the commutation relations
\begin{equation}\label{PLalgebra}
    [W^\mu,P^\nu]=0,\qquad [W^\mu,W^\nu]=-\epsilon^{\mu\nu\alpha\beta}W_\alpha P_\beta
\end{equation}
in the convention $\epsilon_{0123}= 1$. For a momentum eigenstate with $\uvec p =\uvec 0$, the action of $\uvec W$ coincides with that of $m \uvec J$. Hence one can define a relativistic spin operator by
\beq\label{SEdef}
    \uvec S_E=\frac{\uvec W}{P^0}=\uvec J+\frac{\uvec K \times \uvec P}{P^0}  =\uvec J - \uvec R_E \times \uvec P,
\eeq
showing that $\uvec S_E$ corresponds to the internal AM about the relativistic center of energy. Note however that, like the external AM $\uvec R_E\times\uvec P$, it is not a genuine AM operator. An explicit calculation gives indeed~\cite{Born:1935ap,Pryce:1948pf}
\begin{equation}
    [S^i_E,S^j_E]=i\epsilon^{ijk}\left(\delta^{kl}-\frac{P^kP^l}{(P^0)^2}\right)S_E^l
\end{equation}
and shows that $\uvec S_E$ does not obey the $su(2)$ algebra, except when acting on a momentum eigenstate with $\uvec p =\uvec 0$.

\subsection{The relativistic center of mass}

As already mentioned, the relativistic center of energy $\uvec R_E$ does not behave as the spatial part of a Lorentz four-vector. This is not a fundamental issue since the center of a system is just a representative point to which one can attach global properties like mass, momentum and spin. It does not need to behave as an actual physical point. As a result, different observers will generally locate the center of energy at different places inside the convex hull of the system~\cite{Moller:1949,Moller:1949bis}. One may then wonder whether it is possible to define a natural relativistic center whose coordinates would transform as part of a Lorentz four-vector.

Clearly, the center of energy defined in the rest frame of the system plays a special role. Since in that frame the energy coincides with the mass of the system, we will refer to it as the relativistic center of mass. At the classical level, if $r^\mu_E|_\text{rest}$ represents the space-time coordinates of the relativistic center of mass in the rest frame, its components in an arbitrary frame will be defined by $r^\mu_M=\Lambda^\mu_{\phantom{\mu}\nu}r^\nu_E|_\text{rest}$, where $\Lambda^\mu_{\phantom{\mu}\nu}$ represents the Lorentz transformation from the rest frame to the moving frame, i.e. such that $p^\mu=\Lambda^\mu_{\phantom{\mu}\nu}p^\nu_\text{rest}$ with $p^\mu_\text{rest}=(m,\uvec 0)$. Transposing this construction to an operator level is not so trivial. First, one needs to clarify what is meant by ``a system at rest'' in a quantum theory. Since $\uvec P$ is an operator, the rest frame will simply mean the frame in which the expectation value\footnote{The precise meaning of the expectation value will be given in Sec.~\ref{sec:ME}.} of the total momentum operator vanishes
\begin{equation}
    \langle \uvec P\rangle|_\text{rest}=\uvec 0.
\end{equation}
Second, since $R^\mu_E$ involves symmetrized products of operators, the above classical requirements are transposed in a quantum theory into requirements on the expectation values, namely
\begin{equation}
    \langle R^\mu_M\rangle=\Lambda^\mu_{\phantom{\mu}\nu}\langle R^\nu_E\rangle|_\text{rest},\qquad \langle P^\mu\rangle=\Lambda^\mu_{\phantom{\mu}\nu}\langle P^\nu\rangle|_\text{rest}.
\end{equation}

A manifestly covariant generalization of Eq.~\eqref{K} is given by~\cite{Pryce:1948pf,Fleming:1965}
\beq
R_\star^\mu =\tau\,\frac{P^\mu}{M}-\frac{P_\nu M^{\nu\mu}}{M^2},
\eeq
where $\tau$ is the proper time, $P_\nu M^{\nu\mu}/M$ is the covariant form of the boost generators defined by a comoving observer, and $M=\sqrt{P^2}$ is the mass operator. In the rest frame, proper time coincides with the time coordinate\footnote{Note that proper time is defined by $\tau=\langle R^\mu_\star\rangle u_\mu$ with $u^\mu=p^\mu/m$,  whereas the time coordinate is defined by $t=\langle R^0_\star\rangle$.} so that one has as required $\langle R^\mu_\star\rangle|_\text{rest}=\langle R^\mu_E\rangle|_\text{rest}$. The only problems are that $[R^i_\star,P^j]=i\left(\delta^{ij}+\frac{P^iP^j}{M^2}\right)$ and $R^0_\star\neq t$, and have to do with the fact that the operator $\uvec R_\star$ represents a position at a fixed proper time $\tau$ and not at a fixed time coordinate $t$. The solution is to use another operator~\cite{Pryce:1948pf,Fleming:1965}
\beq\label{RM}
R_M^\mu =\left(t+\frac{\uvec P\cdot\uvec K}{M^2}\right)\frac{P^\mu}{P^0}-\frac{P_\nu M^{\nu\mu}}{M^2}
\eeq
that satisfies $[R^i_M,P^j]=i\delta^{ij}$ and $R^0_M=t$. Despite its lack of manifest covariance, it describes in any frame the same world line as $R^\mu_\star$, namely
\begin{equation}
    \{\langle R^\mu_M\rangle\,|\,t\in\text{Reals}\}=\{\langle R^\mu_\star\rangle\,|\,\tau\in\text{Reals}\}.
\end{equation}
In other words, the expectation value $\langle R^\mu_M\rangle$ transforms as a Lorentz four-vector. Moreover, one has obviously $\langle R^\mu_M\rangle|_\text{rest}=\langle R^\mu_E\rangle|_\text{rest}$. 

Combining Eqs.~\eqref{RM} and~\eqref{K}, one finds that the shift between the relativistic centers of energy and mass is given by
\beq
\uvec R_E-\uvec R_M=\frac{\uvec P\times\uvec W}{P^0M^2}.
\eeq
The internal AM operator about the relativistic center of mass $\uvec S_M=\uvec J-\uvec R_M \times \uvec P$ can then be expressed in terms of the Pauli-Luba\'nski pseudovector as follows
\beq
\uvec S_M=\frac{P^0\uvec W-\uvec P W^0}{M^2}.
\eeq
Under Lorentz transformations, it behaves in the same way as the total AM $\uvec J$, i.e. as the axial-vector part $S^i_M=\frac{1}{2}\,\epsilon^{ijk}S^{jk}_M$ of a rank-two antisymmetric Lorentz tensor $S^{\mu\nu}_M=-\epsilon^{\mu\nu\rho\sigma}W_\rho P_\sigma/M^2$. Note however that, like $\uvec R_E$, the components of $\uvec R_M$ do not commute with each other. It is also easy to check that $\uvec S_M$ does not satisfy the $su(2)$ algebra~\cite{Pryce:1948pf} 
\begin{equation}
    [S^i_M,S^j_M]=i\epsilon^{ijk}\left(\delta^{kl}+\frac{P^kP^l}{M^2}\right)S_M^l
\end{equation}
and cannot be considered as a genuine AM operator, except when acting on a momentum eigenstate with $\uvec p =\uvec 0$.

\subsection{The relativistic center of spin}

As stressed above, neither the components of $\uvec R_E$ nor those of $\uvec R_M$ commute with each other~\cite{Born:1935ap,Pryce:1948pf,Moller:1949,Fleming:1965}. The corresponding spin operators $\uvec S_E$ and $\uvec S_M$ do not satisfy the $su(2)$ algebra of AM operators, and hence cannot be considered as generators of rotations. Remarkably, it has been observed that a position operator with commuting components can be obtained by an appropriate weighted average of $R^\mu_E$ and $R^\mu_M$~\cite{Pryce:1935ibt,Pryce:1948pf,Moller:1949}
\begin{equation}\label{RNW}
    R^\mu_c=\frac{P^0 R^\mu_E+MR^\mu_M}{P^0+M},
\end{equation}
proved to be unique by Newton and Wigner~\cite{Newton:1949cq}. Clearly, like $\uvec R_E$ and $\uvec R_M$, the operator $\uvec R_c$ satisfies the commutations relations $[R^i_c,P^j]=i\delta^{ij}$ and $[R^i_c,J^j]=i\epsilon^{ijk}R^k_c$, and therefore qualifies as a position operator.

The corresponding internal AM operator $\uvec S_c=\uvec J-\uvec R_c\times\uvec P$ reads~\cite{Pryce:1935ibt,Pryce:1948pf,Moller:1949,Bogolyubov:1975ps,Choi:2014nea}
\beq
\uvec S_c=\frac{P^0\uvec S_E+M\uvec S_M}{P^0+M}=\frac{\uvec W}{M}-\frac{\uvec P W^0}{M(P^0+M)}.
\eeq
It satisfies the familiar commutation relations
\beq
[S^i_c,S^j_c]=i\epsilon^{ijk}S^k_c,\qquad [S^i_c,R^j_c]=0,\qquad [S^i_c,P^j]=0,
\eeq
and can therefore be interpreted as the generator of internal rotations. For these reasons, the point located at $\langle\uvec R_c\rangle$ has been called the relativistic center of spin in Refs.~\cite{Fleming:1965,Schwartz:2020lys} and the canonical reference point in Ref.~\cite{Lorce:2018zpf}.

Since $P\cdot W=0$, the canonical spin operator can be put in the form~\cite{Keister:1991sb}
\beq\label{covSNW}
S^\mu_c=(L^{-1}_c(P))^\mu_{\phantom{\mu}\nu}\,\frac{W^\nu}{M}
\eeq
with $S^0_c=0$. The operator
\beq
(L^{-1}_c(P))^\mu_{\phantom{\mu}\nu}=\delta^\mu_\nu-\frac{(P^\mu+M\delta^\mu_0)(P_\nu+M\delta^0_\nu)}{M(P^0+M)}+2\frac{\delta^\mu_0P_\nu}{M}
\eeq
satisfies the properties
\begin{equation}
    (L^{-1}_c(P))^\mu_{\phantom{\mu}\nu}P^\nu=M\delta^\mu_0,\qquad P_\nu=M\delta^0_\mu\,(L^{-1}_c(P))^\mu_{\phantom{\mu}\nu}=(L_c(P))_\nu^{\phantom{\nu}\mu}M\delta^0_\mu,
\end{equation}
and can be understood as the operator analog of $(L^{-1}_c(p))^\mu_{\phantom{\mu}\nu}$, the canonical (rotationless) four-vector boost from the moving frame to the rest frame~\cite{Krause:1977yr,Fahnline:1982}. Sandwiching Eq.~\eqref{covSNW} between momentum eigenstates, one gets 
\begin{equation}
    \begin{aligned}
    \langle p|S^i_c|p\rangle&=\frac{1}{m}\,(L^{-1}_c(p))^i_{\phantom{i}\nu}\langle p|W^\nu|p\rangle\\
    &=\frac{1}{m}\,\langle p|U^{-1}(L^{-1}_c(p))W^iU(L^{-1}_c(p))|p\rangle\\
    &=\langle p_\text{rest}|J^i|p_\text{rest}\rangle.
    \end{aligned}
\end{equation}
Since momentum eigenstates are covariantly normalized, one can write $\langle p|p\rangle=\langle p_\text{rest}|p_\text{rest}\rangle$ and therefore conclude that the expectation value of canonical spin in an arbitrary frame gives the same value as the rest-frame AM 
\begin{equation}
    \frac{\langle p|S^i_c|p\rangle}{\langle p|p\rangle}=\frac{\langle p_\text{rest}|J^i|p_\text{rest}\rangle}{\langle p_\text{rest}|p_\text{rest}\rangle}.
\end{equation}
In particular, a spin-$1/2$ momentum eigenstate with canonical polarization described by the unit vector $\uvec s$ is an eigenstate of the canonical spin component along $\uvec s$
\begin{equation}
    \uvec s\cdot\uvec S_c|p,\uvec s\rangle=-\frac{\ucal S_\mu W^\mu}{m}|p,\uvec s\rangle=\tfrac{1}{2}\,|p,\uvec s\rangle, 
\end{equation}
where the (canonical) covariant polarization vector is defined as
\beq\label{covpol}
\ucal S^\mu(p,\uvec s)=(L_c(p))^\mu_{\phantom{\mu}\nu}\ucal S^\nu_\text{rest}=\left(\frac{\uvec p\cdot\uvec s}{m},\uvec s+\frac{\uvec p(\uvec p\cdot\uvec s)}{m(p^0+m)}\right)
\eeq
with $\ucal S^\mu_\text{rest}=(0,\uvec s)$.

As a final remark, it is interesting to express the generators of boosts in terms of the canonical position and spin operators
\beq\label{KRc}
\uvec K=\uvec P t-P^0\uvec R_c-\frac{\uvec P\times\uvec S_c}{P^0+M}.
\eeq
This relation clearly shows that the operators $\uvec K$ generate not only boosts of the system (first two terms), but also rotations in internal or spin space (last term) known as the Wigner rotations~\cite{Wigner:1939cj,Polyzou:2012ut}, which are responsible for the non-commutativity of the components of $\uvec K$.

\section{Expectation values}\label{sec:ME}

Now that the notion of spin and the question of the pivot have been reviewed, we turn our attention to the computation of the expectation values of the Lorentz generators. They require particular care because matrix elements of the type\footnote{The operator $O(r)$ is required not to involve any explicit factor of $r^i$.} $\la p,\uvec s|\int\ud^3r\,r^i O(r)|p,\uvec s\ra$ are usually ill-defined, being either infinite or, by symmetry, zero. We present in the following the two approaches designed to deal with this problem.

We will always work with covariantly normalized momentum eigenstates $\langle p_f,\uvec s|p_i,\uvec s\rangle=(2\pi)^3\,2p^0_i\,\delta^{(3)}(\uvec p_f-\uvec p_i)$, so that the completeness relation (resolution of the identity) becomes
\beq \label{compl} 
I = \sum_{\sigma=\pm}\int\frac{\ud^3p}{(2\pi)^3 2p_0}\, |p,\sigma\uvec s\rangle \langle p, \sigma\uvec s |. 
\eeq
For later convenience, we introduce the average four-momentum and four-momentum transfer variables
\begin{equation}
    \bar p=\tfrac{1}{2}(p_f+p_i),\qquad \Delta=p_f-p_i
\end{equation}
which obey the constraints
\begin{equation}\label{onshell}
\bar p\cdot\Delta=0,\qquad \bar p^2=m^2-\frac{\Delta^2}{4}
\end{equation}
owing to the onshell relations $p^2_i=p^2_f=m^2$.

\subsection{Standard approach}

The standard approach is a prescription which consists in considering first the matrix element with nonzero momentum transfer $\uvec\Delta$, and then taking the forward limit $\uvec\Delta\to \uvec 0$ at the end of the calculation~\cite{Jaffe:1989jz,Shore:1999be,Bakker:2004ib,Leader:2013jra,Lowdon:2017idv,Cotogno:2019xcl,Lorce:2019sbq}
\begin{equation}\label{standard}
\la p,\uvec s|\int\ud^3r\,r^i O(r)| p,\uvec s\ra\equiv\lim_{\uvec\Delta\to\uvec 0}\la \bar p+\tfrac{\Delta}{2},\uvec s|\int\ud^3r\,r^i O(r)|\bar p-\tfrac{\Delta}{2},\uvec s\ra.
\end{equation}
Note that $\bar{\uvec p}$ is kept fixed so that $\bar{\uvec p}=\uvec p$, and from Eq.~\eqref{onshell}
\begin{equation}
    \bar p^0=\frac{1}{2}\left(\sqrt{(\uvec p+\tfrac{\uvec\Delta}{2})^2+m^2}+\sqrt{(\uvec p-\tfrac{\uvec\Delta}{2})^2+m^2}\right)
\end{equation}
whereas $p^0=\sqrt{\uvec p^2+m^2}$.

For convenience, we will also work at $t=0$ in the following. One can write using translation symmetry and the Leibniz rule\footnote{Instead of the Leibniz rule, one can use directly the distributional identity~\cite{Strichartz:1994} $f(x)\nabla\delta(x)=f(0)\nabla\delta(x)-\left[\nabla f(x)\right]_{x=0}\delta(x)$.}
\begin{equation}\label{EVder}
\begin{aligned}
\la \bar p+\tfrac{\Delta}{2},\uvec s|&\int\ud^3r\,r^i O(r)|\bar p-\tfrac{\Delta}{2},\uvec s\ra\\
&=\int\ud^3r\,e^{-i\uvec\Delta\cdot\uvec r}r^i \la \bar p+\tfrac{\Delta}{2},\uvec s|O(0)|\bar p-\tfrac{\Delta}{2},\uvec s\ra\\
&=\left[(2\pi)^3i\nabla^i\delta^{(3)}(\uvec\Delta)\right]\la \bar p+\tfrac{\Delta}{2},\uvec s|O(0)|\bar p-\tfrac{\Delta}{2},\uvec s\ra\\
&=i\nabla^i\!\left[(2\pi)^3\delta^{(3)}(\uvec\Delta)\,\la \bar p+\tfrac{\Delta}{2},\uvec s|O(0)|\bar p-\tfrac{\Delta}{2},\uvec s\ra\right]-\left[i\nabla^i\la \bar p+\tfrac{\Delta}{2},\uvec s|O(0)|\bar p-\tfrac{\Delta}{2},\uvec s\ra\right](2\pi)^3\delta^{(3)}(\uvec\Delta)\\
&=\la p,\uvec s|O(0)|p,\uvec s\ra\,(2\pi)^3i\nabla^i\delta^{(3)}(\uvec\Delta)+\left[-i\nabla^i\la \bar p+\tfrac{\Delta}{2},\uvec s|O(0)|\bar p-\tfrac{\Delta}{2},\uvec s\ra\right]_{\uvec\Delta=\uvec 0}(2\pi)^3\delta^{(3)}(\uvec\Delta),
\end{aligned}
\end{equation}
where $\nabla^i=\frac{\partial}{\partial\Delta^i}$. One then finds for the expectation value
\begin{equation}\label{fullEV}
\frac{\la p,\uvec s|\int\ud^3r\,r^i O(r)|p,\uvec s\ra}{\la p,\uvec s|p,\uvec s\ra}=\frac{\la p,\uvec s|O(0)|p,\uvec s\ra}{\la p,\uvec s|p,\uvec s\ra}\,(2\pi)^3i\nabla^i\delta^{(3)}(\uvec 0)+\frac{1}{2p^0}\left[-i\nabla^i\la \bar p+\tfrac{\Delta}{2},\uvec s|O(0)|\bar p-\tfrac{\Delta}{2},\uvec s\ra\right]_{\uvec\Delta=\uvec 0}.
\end{equation}
Note that in the second term one should not forget that both $\Delta^0$ and $\bar p^0$ are understood as functions of $\uvec\Delta$, derived from the onshell constraints~\eqref{onshell}. 

In order to make sense of the highly divergent term in Eq.~\eqref{fullEV}, one has to work with normalizable states and hence introduce some wave packet. Calculations with explicit forms for the wave packets are pretty lengthy and cumbersome, but they show that the highly divergent term corresponds actually to the contribution from the center of the wave packet~\cite{Jaffe:1989jz,Bakker:2004ib}. Since one is interested in the internal structure of the nucleon and not in the details of the wave packet, this term is usually discarded.

\subsection{Phase-space approach}

Discarding the highly divergent term in Eq.~\eqref{fullEV} is tantamount to throwing the baby out with the bathwater. Indeed, even though we are not interested in the details of the wave packet, it is important to retain the information about the position of its center. This can easily be achieved within the phase-space approach developed in Refs.~\cite{Lorce:2018zpf,Lorce:2017wkb,Lorce:2018egm}, and used in Ref.~\cite{Lorce:2020onh} to elucidate the momentum dependence of relativistic charge distributions. 	

In a quantum theory, states localized around the position $\uvec r$ at time $t=0$ are defined as the following Fourier transform of momentum eigenstates
\begin{equation}\label{pos}
|\uvec r\rangle\equiv\int\frac{\ud^3p}{(2\pi)^3\sqrt{2p^0}}\,e^{-i\uvec p\cdot\uvec r}\,|p\rangle
\end{equation}
and are normalized as $\langle\uvec r_f|\uvec r_i\rangle=\delta^{(3)}(\uvec r_f-\uvec r_i)$. Wave packets associated with physical states normalized as $\langle\psi|\psi\rangle=1$ are then defined by
\begin{equation}
\tilde\psi(\uvec p)\equiv\frac{\langle p|\psi\rangle}{\sqrt{2p^0}}\qquad\text{so that}\qquad
\psi(\uvec r)\equiv\langle\uvec r|\psi\rangle=\int\frac{\ud^3p}{(2\pi)^3}\,e^{i\uvec p\cdot\uvec r}\,\tilde\psi(\uvec p).
\end{equation}
A physical state cannot be simultaneously in an eigenstate of both position and momentum. But it was shown long ago how to define a quantity, analogous to a density matrix, which gives the quantum weight that a system in a definite state $|\psi\ra$ will be found to have average position $\uvec{\ucal R}$ and average momentum $\uvec{p}$. This quantum phase-space or Wigner distribution is given by~\cite{Wigner:1932eb,Hillery:1983ms}
 \begin{equation} \label{Wigner}
\begin{aligned}
\rho_\psi(\uvec{\ucal R},\uvec p)&\equiv\int\ud^3z\,e^{-i\uvec p\cdot\uvec z}\,\psi^*(\uvec{\ucal R}-\tfrac{\uvec z}{2})\psi(\uvec{\ucal R}+\tfrac{\uvec z}{2})\\
&=\int\frac{\ud^3q}{(2\pi)^3}\,e^{-i\uvec q\cdot\uvec{\ucal R}}\,\tilde\psi^*(\uvec p+\tfrac{\uvec q}{2})\tilde\psi(\uvec p-\tfrac{\uvec q}{2})
\end{aligned}
\end{equation}
and is not positive-definite because of the Heisenberg uncertainty relations. It is thus a \emph{quasi}-probabilistic density matrix. It can be seen that $\uvec{\ucal R}$ is the average of the initial and final position vectors and $\uvec p$ is the average of the initial and final momenta, respectively, in Eq.~\eqref{Wigner}. One recovers a probabilistic interpretation after integration over either average position or average momentum
\begin{equation}\label{Rprob}
\int\ud^3\ucal R\,\rho_\psi(\uvec{\ucal R},\uvec p)=|\tilde\psi(\uvec p)|^2,\qquad
\int\frac{\ud^3 p}{(2\pi)^3}\,\rho_\psi(\uvec{\ucal R},\uvec p)=|\psi(\uvec{\ucal R})|^2.
\end{equation}

One can now express the expectation of any operator $O$ as a phase-space integral~\cite{Wigner:1932eb,Hillery:1983ms} over $\uvec{\ucal R}$ and $\uvec p$ of the Wigner distribution times the expectation value of $O$ in a state having average position $\uvec{\ucal R}$ and average momentum  $\uvec p$, namely
   \begin{equation}\label{PSampl}
\langle\psi| O|\psi\rangle=\int\frac{\ud^3 p}{(2\pi)^3}\,\ud^3\ucal R\,\rho_\psi(\uvec{\ucal R},\uvec p)\,\langle O\rangle_{\uvec{\ucal R},\uvec p},
\end{equation}
where\footnote{We note in passing that the covariant formulation of the phase-space approach proposed in Ref.~\cite{Lorce:2018egm} is not totally rigorous in the implementation of the onshell constraints since $\delta(p^2_f-m^2)\,\delta(p^2_i-m^2)\neq\delta(2\bar p\cdot\Delta)\,\delta(\bar p^2+\frac{\Delta^2}{4}-m^2)$, leading to a factor $2\bar p^0$ in the denominator of Eq.~\eqref{intampl} instead of $\sqrt{2p^0_f}\sqrt{2p^0_i}=\sqrt{4(\bar p^0)^2-(\Delta^0)^2}$. In the context of Ref.~\cite{Lorce:2018egm} this did not matter since the results were derived in Lorentz frames where $\Delta^0=0$.}
\begin{equation}\label{intampl}
\langle O\rangle_{\uvec{\ucal R},\uvec p}=\int\frac{\ud^3\Delta}{(2\pi)^3}\,e^{i\uvec\Delta\cdot\uvec{\ucal R}}\,\frac{\langle\bar p+\tfrac{\Delta}{2}| O|\bar p-\tfrac{ \Delta}{2}\rangle}{\sqrt{4\left(\bar p^0\right)^2-\left(\Delta^0\right)^2}} \qquad \textrm{with} \qquad  \bar{\uvec{p}} = \uvec{p}.
\end{equation}
Although originally developed in the non-relativistic context, this formalism carries over to Quantum Field Theory~\cite{BialynickiBirula:1991tx}. Requiring the localization\footnote{In Quantum Field Theory, it is often stated that a particle cannot be localized over distances smaller than the Compton wavelength. It does not mean that one cannot define a position eigenstate $|\uvec r\rangle$ for a system, but it just expresses the limit of the single onshell particle picture. Note also that if the average momentum is non-zero, then the bound is determined by $1/\langle p^0\rangle$ and not $1/m$~\cite{Burkardt:2000za}.} of a system simultaneously in the three spatial dimensions leads automatically to a position operator with mutually commuting components, i.e. to $\uvec R_c$ which, as mentioned, is the only relativistic operator with that property~\cite{Pryce:1948pf,Newton:1949cq,Foldy:1949wa}. It has indeed been shown in Refs.~\cite{Newton:1949cq,Pavsic:2017orp} that the states $|\uvec r\rangle$ defined by Eq.~\eqref{pos} are eigenstates of the canonical position operator
\begin{equation}
    \uvec R_c|\uvec r\rangle=\uvec r|\uvec r\rangle.
\end{equation}
Since wave-packet details are encoded in $\rho_\psi(\uvec{\ucal R},\uvec p)$ in Eq.~\eqref{PSampl}, one can interpret $\langle O\rangle_{\uvec{\ucal R},\uvec p}$ as the part associated with the internal structure of the system localized in the Wigner sense around the (average) canonical position $\langle\uvec R_c\rangle_{\uvec{\ucal R},\uvec p}=\uvec{\ucal R}$ with (average) momentum $\langle\uvec P\rangle_{\uvec{\ucal R},\uvec p}=\uvec p$. As noted in Ref.~\cite{Lorce:2018zpf}, the standard expectation value~\eqref{standard} corresponds simply to an average over $\uvec{\ucal R}$ of the phase-space expectation value~\eqref{intampl}
\begin{equation}\label{Raverage}
\int\frac{\ud^3\ucal R}{(2\pi)^3\delta^{(3)}(\uvec 0)}\,\langle O\rangle_{\uvec{\ucal R},\uvec p}=\int\ud^3\Delta\,\delta^{(3)}(\uvec\Delta)\,\frac{\langle\bar p+\tfrac{\Delta}{2}| O|\bar p-\tfrac{ \Delta}{2}\rangle}{(2\pi)^32 p^0\delta^{(3)}(\uvec 0)}=\frac{\langle p|O|p\rangle}{\langle p|p\rangle}.
\end{equation} 

Proceeding similarly to Eq.~\eqref{EVder}, one finds including now polarization~\cite{Lorce:2018zpf}
\begin{equation}\label{PSEV}
\langle\int\ud^3 r\,r^iO(r)\rangle_{\uvec{\ucal R},\uvec p,\uvec s}=\ucal R^i\,\frac{\la p,\uvec s|O(0)|p,\uvec s\ra}{2p^0}+\frac{1}{2p^0}\left[-i\nabla^i\la \bar p+\tfrac{\Delta}{2},\uvec s|O(0)|\bar p-\tfrac{\Delta}{2},\uvec s\ra\right]_{\uvec\Delta=\uvec 0}.
\end{equation}
Comparing with Eq.~\eqref{fullEV}, one observes that the highly divergent term is replaced by a well-defined expression in the phase-space approach, which confirms its interpretation as the contribution associated with the center of the wave packet. According to Eq.~\eqref{Raverage} one should recover Eq.~\eqref{fullEV} from an average of Eq.~\eqref{PSEV} over $\uvec{\ucal R}$. Indeed, the second term in Eq.~\eqref{PSEV} is independent of $\uvec{\ucal R}$ and is therefore not affected by the average over $\uvec{\ucal R}$. The first term however is linear in $\uvec{\ucal R}$ and gives the ambiguous contribution in Eq.~\eqref{fullEV}, since one can formally write $\int\ud^3\ucal R\,\ucal R^i=\lim_{\uvec\Delta\to\uvec 0}\int\ud^3\ucal R\,e^{-i\uvec\Delta\cdot\uvec{\ucal R}}\ucal R^i=(2\pi)^3\,i\nabla^i\delta^{(3)}(\uvec 0)$.

As mentioned earlier, in the literature one is usually interested in the internal part and accordingly considers only the second term in Eq.~\eqref{PSEV} or Eq.~\eqref{fullEV}~\cite{Jaffe:1989jz,Shore:1999be,Bakker:2004ib,Leader:2013jra}. This can be justified by choosing the (average) relativistic center of spin as the origin of the coordinate system, i.e. by setting $\uvec{\ucal R}=\uvec 0$
\begin{equation}\label{standardform}
\langle\int\ud^3 r\,r^iO(r)\rangle_{\uvec 0,\uvec p,\uvec s}=\frac{1}{2p^0}\left[-i\nabla^i\la \bar p+\tfrac{\Delta}{2},\uvec s|O(0)|\bar p-\tfrac{\Delta}{2},\uvec s\ra\right]_{\uvec\Delta=\uvec 0}.
\end{equation}
The drawback is that when we change the Lorentz frame, a shift of the origin is required for the new origin to coincide with the relativistic center of spin in the new Lorentz frame, because the coordinates of the latter do not transform as part of a four-vector. In fact, we can obtain the internal part independently of the choice for the origin as follows
\begin{equation}\label{genform}
\langle\int\ud^3 r\,(r^i-\ucal R^i)\,O(r)\rangle_{\uvec{\ucal R},\uvec p,\uvec s}=\frac{1}{2p^0}\left[-i\nabla^i\la \bar p+\tfrac{\Delta}{2},\uvec s|O(0)|\bar p-\tfrac{\Delta}{2},\uvec s\ra\right]_{\uvec\Delta=\uvec 0}
\end{equation}
using the relation $\langle\int\ud^3 r\,O(r)\rangle_{\uvec{\ucal R},\uvec p,\uvec s}=\la p,\uvec s|O(0)|p,\uvec s\ra/2p^0$.

\section{AM sum rules}\label{sec:AMSR}

All the spin sum rules are based on a decomposition of the QCD EMT into quark and gluon parts
\begin{align}
T^{\mu\nu}(r)&=T^{\mu\nu}_q(r)+T^{\mu\nu}_g(r).
\end{align}
In the Belinfante form, they are given by 
\begin{equation}
\begin{aligned}
T^{\mu\nu}_q(r)&=\barpsi(r)\gamma^{\{\mu}\tfrac{i}{2}\LRD^{\nu\}}\psi(r),\\
T^{\mu\nu}_g(r)&=-F^{\mu\lambda}(r)F^{\nu}_{\phantom{\nu}\lambda}(r)+\frac{1}{4}\,g^{\mu\nu}F^{\alpha\beta}(r)F_{\alpha\beta}(r)
\end{aligned}
\end{equation}
with $\LRD_\mu=\overset{\rightarrow}{\partial}\!\!\!\!\phantom{\partial}_\mu-\overset{\leftarrow}{\partial}\!\!\!\!\phantom{\partial}_\mu-2igA_\mu(r)$ the symmetric covariant derivative, and $a^{\{\mu}b^{\nu\}}=\frac{1}{2}(a^\mu b^\nu+a^\nu b^\mu)$. The corresponding decompositions of the other tensors follow automatically. 

For a spin-$1/2$ state with mass $m$, the matrix elements of the Belinfante EMT can be parametrized in general as~\cite{Kobzarev:1962wt,Pagels:1966zza,Ji:1996ek}
\begin{subequations}\label{param}
\beq
\la p_f,\uvec s_f|T^{\mu\nu}_a(0)|p_i,\uvec s_i\ra=\baru(p_f,\uvec s_f)\Gamma^{\mu\nu}_a(\bar p,\Delta)u(p_i,\uvec s_i),\qquad\qquad a=q,g
\eeq
with
\beq
\Gamma^{\mu\nu}_a(\bar p,\Delta)=\frac{\bar p^{\{\mu}\gamma^{\nu\}}}{m}\,A_a(\Delta^2)+\frac{\bar p^{\{\mu}i\sigma^{\nu\}\rho}\Delta_\rho}{2m}\,B_a(\Delta^2)+\frac{\Delta^\mu\Delta^\nu-g^{\mu\nu}\Delta^2}{m}\,C_a(\Delta^2)+mg^{\mu\nu}\bar C_a(\Delta^2).
\eeq
\end{subequations}
Here $\uvec s_i$ ($\uvec s_f$) is the unit rest-frame polarization vector associated with the initial (final) state. The EMT form factors $A_a,B_a,C_a,\bar C_a$ are Lorentz-invariant functions of $\Delta^2$, which is the only independent Lorentz scalar formed with four-momenta as indicated by the onshell constraints~\eqref{onshell}.

Using this parametrization, the standard expectation value of the quark and gluon parts of the four-momentum operator reads~\cite{Ji:1997pf}
\begin{equation}\label{EVmom}
\frac{\la p,\uvec s| P^\mu_a |p,\uvec s\ra}{\la p,\uvec s|p,\uvec s\ra}=p^\mu A_a(0)+\frac{m^2}{p^0}\,g^{\mu 0}\bar C_a(0).
\end{equation}
The same expression is obtained using the phase-space expectation value $\langle P^\mu_a\rangle_{\uvec{\ucal R},\uvec p,\uvec s}$, in agreement with Eq.~\eqref{Raverage}. Four-momentum conservation then implies two constraints
\begin{equation}\label{momSR}
\sum_{a}A_a(0)=1, \qquad \sum_{a}\bar C_a(0)=0.
\end{equation}
Note that the constraint for $\bar C$ holds also when $\Delta^2\neq 0$ as a consequence of the conservation of $T^{\mu\nu}$.

Even though $|p,\uvec s\rangle$ is in general not an eigenstate of the AM operator\footnote{Surprisingly, it is stated in Refs.~\cite{Ji:2020ena,Ji:2020hii} that transverse AM does not commute with the QCD Hamiltonian. This is obviously incorrect for otherwise the Hamiltonian would not be invariant under rotations, breaking therefore Poincar\'e symmetry. The actual reason is that transverse AM does not commute with the three-momentum operator $[P^i,J^j]=i\epsilon^{ijk}P^k$.}, the expectation value understood in the sense of Eq.~\eqref{PSEV} is well-defined. Performing an expansion in $\Delta$ of the parametrization~\eqref{param} for the case $\uvec s_f=\uvec s_i=\uvec s$, one gets
\begin{equation}
\begin{aligned}
\la \bar p+\tfrac{\Delta}{2},\uvec s|T^{\mu\nu}_a(0)|\bar p-\tfrac{\Delta}{2},\uvec s\ra&=2\left[p^\mu p^\nu A_a(0)+m^2g^{\mu\nu}\bar C_a(0)\right]\\
&\quad+\frac{i\Delta_\lambda}{m}\left[\left(p^\mu\epsilon^{\lambda\nu\alpha\beta}+p^\nu\epsilon^{\lambda\mu\alpha\beta}\right)\frac{A_a(0)+B_a(0)}{2}\right.\\
&\quad+\left.\epsilon^{0\lambda\alpha\beta}\,\frac{p^\mu p^\nu A_a(0)+m^2g^{\mu\nu}\bar C_a(0)}{p^0+m}\right]\ucal S_\alpha p_\beta+\ucal O(\Delta^2),
\end{aligned}
\end{equation}
where we remind that $p^\mu=(\sqrt{\uvec p^2+m^2},\uvec p)$ with $\uvec p=\bar{\uvec p}$, and $\ucal S^\mu(p,\uvec s)$ is the covariant polarization vector defined in Eq.~\eqref{covpol}. We then obtain from Eq.~\eqref{PSEV} our master formula
\beq\label{master}
\begin{aligned}
\langle\int\ud^3 r\,r^iT^{\mu\nu}_a(r)\rangle_{\uvec{\ucal R},\uvec p,\uvec s}&=\ucal R^i\,\frac{p^\mu p^\nu A_a(0)+m^2g^{\mu\nu}\bar C_a(0)}{p^0}\\
&\quad+(\delta^i_\lambda-\delta^0_\lambda\,\tfrac{p^i}{p^0})\,\frac{p^{\{\mu}\epsilon^{\nu\}\lambda\alpha\beta}\ucal S_\alpha p_\beta}{p^0m}\,\frac{A_a(0)+B_a(0)}{2}\\
&\quad-\frac{\epsilon^{0i\alpha\beta}\ucal S_\alpha p_\beta}{2m(p^0+m)}\,\frac{p^\mu p^\nu A_a(0)+m^2g^{\mu\nu}\bar C_a(0)}{p^0}.
\end{aligned}
\eeq
The first line corresponds to the contribution from the motion of the center of the wave packet, the second line gives the tensorial part of the canonical internal contribution, and the last line gives the non-tensorial part arising due to the Wigner rotation~\cite{Bakker:2004ib,Lorce:2017isp}. As stressed earlier, in deriving this formula it is important to keep in mind that $\Delta^0=\bar{\uvec p}\cdot\uvec\Delta/\bar p^0$ when differentiating w.r.t. $\Delta^i$.

\subsection{Relativistic positions and spins}

At $t=0$, the Poincar\'e generators read
\beq
K^i=-\int\ud^3r\,r^iT^{00}(r), \qquad J^i=\epsilon^{ijk}\int\ud^3r\,r^jT^{0k}(r).
\eeq
Using our master formula~\eqref{master} in the rest frame defined by $\uvec p=\uvec 0$, we get  
\beq
    \langle\uvec K\rangle_{\uvec{\ucal R},\uvec 0,\uvec s}=-m\uvec{\ucal R}\sum_a\left[A_a(0)+\bar C_a(0)\right],\qquad \langle\uvec J\rangle_{\uvec{\ucal R},\uvec 0,\uvec s}=\frac{\uvec s}{2}\,\sum_a\left[A_a(0)+B_a(0)\right].
\eeq
It then follows from the constraints~\eqref{momSR} derived from four-momentum conservation that $\langle\uvec K\rangle_{\uvec{\ucal R},\uvec 0,\uvec s}=-m\uvec{\ucal R}$. Moreover, by definition of the rest-frame AM we must have $\langle\uvec J\rangle_{\uvec{\ucal R},\uvec 0,\uvec s}=\tfrac{1}{2}\uvec s$, so that AM conservation implies the additional constraint~\cite{Kobzarev:1962wt,Ji:1997pf,Bakker:2004ib} 
\beq\label{AMSR}
\sum_a\left[A_a(0)+B_a(0)\right]=1,
\eeq
and hence the absence of anomalous gravitomagnetic moment $\sum_aB_a(0)=0$~\cite{Teryaev:1999su,Brodsky:2000ii} when combined with Eq.~\eqref{momSR}. The same constraints can alternatively be obtained from the requirement that $\uvec J$ and $\uvec K$ be the generators of Lorentz transformations in any frame~\cite{Bakker:2004ib,Lowdon:2017idv,Cotogno:2019xcl,Lorce:2019sbq}.

\begin{figure}[h]
    \centering
    \includegraphics[scale=.65]{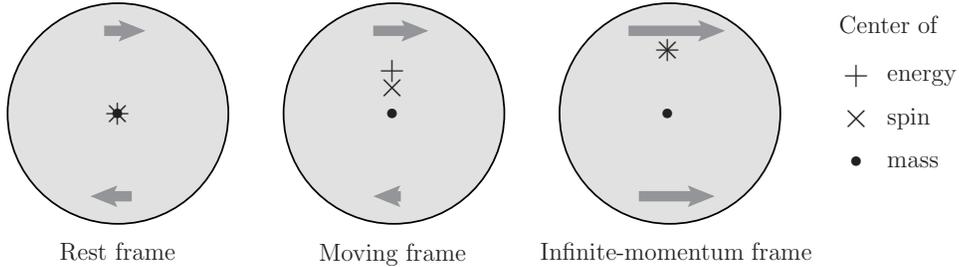}
    \caption{Illustration of the relative positions of the relativistic centers of energy, mass and spin inside a transversely polarized nucleon (the arrows indicate the local momentum seen by the observer). For simplicity, Lorentz contraction effects are not represented. In the rest frame of the nucleon, the three relativistic centers coincide. In a general moving frame they all differ. In the infinite-momentum frame, the relativistic centers of energy and spin merge and are found half a Compton wavelength away from the relativistic center of mass.}
    \label{fig:CoM}
\end{figure}

Expectation values in an arbitrary frame, where $\uvec p\neq\uvec 0$, will always be understood in the sense of Eq.~\eqref{intampl}. For convenience we will drop from now on the explicit phase-space labels $\uvec{\ucal R},\uvec p,\uvec s$. We find~\cite{Lorce:2018zpf}
\begin{equation}
    \langle\uvec K\rangle=-p^0\uvec{\ucal R}-\frac{\uvec p\times\uvec s}{2(p^0+m)},\qquad
    \langle\uvec J\rangle=\uvec{\ucal R}\times\uvec p+\tfrac{1}{2}\uvec s,
\end{equation}
Since we work with four-momentum eigenstates and implicitly symmetrized products, we can write
\begin{equation}\label{idP}
    \langle P^\mu \uvec K\rangle\equiv\langle \tfrac{1}{2}(P^\mu \uvec K+\uvec K P^\mu)\rangle =\langle\bar p^\mu\uvec K\rangle=p^\mu\langle\uvec K\rangle,
\end{equation}
and similarly for $\langle P^\mu\uvec J\rangle$. We recover therefore the well-known result that the expectation value of the Pauli-Luba\'nski pseudovector is proportional to the covariant polarization vector defined in Eq.~\eqref{covpol}
\begin{equation}
    \langle W^\mu\rangle=\frac{m}{2}\,\ucal S^\mu(p,\uvec s).
\end{equation}
Finally, we can express the external contribution for any choice of position vector $\uvec R$ as
\begin{equation}\label{id}
    \langle \uvec R\times\uvec P\rangle=\langle \uvec R\rangle\times\uvec p=\langle \uvec R\rangle\times\langle\uvec P\rangle=\langle \langle\uvec R\rangle\times\uvec P\rangle.
\end{equation}
The expectation values of the three sets of position and spin operators discussed in Sec.~\ref{sec:spin} can then easily be computed. The results are collected in Table~\ref{tab1} and the relative positions of the three relativistic centers are represented in Fig.~\ref{fig:CoM}.
\begin{table}[th!]
\begin{center}
\caption{\footnotesize{Expectation values of the main three position operators and of the corresponding spin operators.}}\label{tab1}
\begin{tabular}{@{\quad}c@{\quad}|@{\quad}c@{\quad}|@{\quad}c@{\quad}}\whline
Relativistic center of&$\langle\uvec R_X\rangle$&$\langle\uvec S_X\rangle=\langle\uvec J\rangle-\langle\uvec R_X\times\uvec P\rangle$\\
\hline
energy ($X=E$)&$\uvec{\ucal R}+\frac{\uvec p\times\uvec s}{2p^0(p^0+m)}$&$\frac{m}{2p^0}\left(\uvec s+\frac{\uvec p(\uvec p\cdot\uvec s)}{m(p^0+m)}\right)$\\
mass ($X=M$)&$\uvec{\ucal R}-\frac{\uvec p\times\uvec s}{2m(p^0+m)}$&$\frac{p^0}{2m}\left(\uvec s-\frac{\uvec p(\uvec p\cdot\uvec s)}{p^0(p^0+m)}\right)$\\
spin ($X=c$)&$\uvec{\ucal R}$&$\tfrac{1}{2}\uvec s$\\
\whline
\end{tabular}
\end{center}
\end{table}

There are many interesting observations to make. First, the explicit calculation confirms that the center of the wave packet is nothing but the relativistic center of spin~\cite{Lorce:2018zpf}, which lies on the segment joining the relativistic centers of mass and energy. In the rest frame defined by $\uvec p=\uvec 0$, the three sets of center and spin coincide. If we decompose three-vectors into longitudinal and transverse components relative to $\uvec p$
\begin{equation}
    \uvec V=(\uvec V\cdot\hat{\uvec p})\,\hat{\uvec p}+\uvec V_{\!\perp}
\end{equation}
with $\hat{\uvec p}=\uvec p/|\uvec p|$, one sees that the longitudinal components of position and spin (i.e. helicity) also coincide in any frame
\begin{equation}
\begin{aligned}
    \langle\uvec R_E\rangle\cdot\hat{\uvec p}&=  \langle\uvec R_M\rangle\cdot\hat{\uvec p}=  \langle\uvec R_c\rangle\cdot\hat{\uvec p}=\uvec{\ucal R}\cdot\hat{\uvec p},\\
   \langle\uvec S_E\rangle\cdot\hat{\uvec p} &=  \langle\uvec S_M\rangle\cdot\hat{\uvec p}=  \langle\uvec S_c\rangle\cdot\hat{\uvec p}=\tfrac{1}{2}\uvec s\cdot\hat{\uvec p}.
\end{aligned}
\end{equation}
Differences appear only in the corresponding transverse components and illustrate why transverse spin is always more complicated than longitudinal spin. In particular, there are three different possible definitions of transverse spin:
\begin{itemize}
\item If one defines spin as the internal AM about the relativistic center of energy, then transverse spin is Lorentz \emph{contracted}
\begin{equation}
    \langle \uvec S_{E\perp}\rangle=\gamma^{-1}\,\tfrac{1}{2}\uvec s_\perp
\end{equation}
with $\gamma=p^0/m$ the Lorentz boost factor. 
\item If one defines instead spin as the internal AM about the relativistic center of mass, then transverse spin is Lorentz \emph{dilated} 
\begin{equation}
    \langle \uvec S_{M\perp}\rangle=\gamma\,\tfrac{1}{2}\uvec s_\perp
\end{equation}
This is expected since in this case internal AM transforms in the same way as total AM, i.e. as the axial-vector part of an antisymmetric rank-two tensor, whose transverse components are known to increase linearly with energy~\cite{Landau:1951}. 
\item Finally, if one defines spin as the internal AM about the relativistic center of spin, then transverse spin is \emph{invariant} under Lorentz boosts 
\begin{equation}
    \langle \uvec S_{c\perp}\rangle=\tfrac{1}{2}\uvec s_\perp
\end{equation}
just like longitudinal spin, which is in agreement with the result found in Ref.~\cite{Bakker:2004ib}.
\end{itemize}
It is seen that the different results for the expectation values of the three kinds of spin operators, in the case of transverse spin, is simply due to the fact that they correspond to different versions of what is meant by the internal AM or ``spin''.

Note that in the infinite-momentum frame $|\uvec p|\to\infty$, the relativistic center of spin merges with the relativistic center of energy and is found half a Compton wavelength away from the relativistic center of mass when the polarization is purely transverse~\cite{Burkardt:2002hr,Burkardt:2005hp,Lorce:2018zpf}
\begin{equation}
    \lim_{|\uvec p|\to\infty}\left(\uvec{\ucal R}-\langle\uvec R_M\rangle\right)=\frac{\hat{\uvec p}\times\uvec s}{2m}=\frac{\hat{\uvec p}\times\uvec s}{2}\,\frac{\hbar}{mc}.
\end{equation}
It may then seem surprising that in this limit $\langle \uvec S_{E\perp}\rangle$ vanishes whereas $\langle \uvec S_{c\perp}\rangle$ remains non-zero for a state with non-zero transverse polarization. The reason is that the difference in the definition of the internal AM is given by
\begin{equation}
    \langle\uvec S_{c}\rangle-\langle\uvec S_{E}\rangle=\left(\langle\uvec R_E\rangle-\langle\uvec R_c\rangle\right)\times\uvec p=\frac{(\uvec p\times\uvec s)\times\uvec p}{2p^0(p^0+m)}.
\end{equation}
In the infinite-momentum frame, the infinitely small transverse shift is multiplied by an infinitely large momentum, leading to a finite difference.

\subsection{Spin sum rules}

We are now ready to derive a generic spin sum rule. By this we mean a decomposition of the expectation value of internal AM into quark and gluon contributions
\begin{equation}
    \langle\uvec S_X\rangle=\sum_a\langle\uvec S_{X,a}\rangle,\qquad X=E,M,c,
\end{equation}
where
\begin{equation}\label{spinop}
    S^i_{X,a}\equiv\int\ud^3r\,\epsilon^{ijk}\left(r^j-\langle R^j_X\rangle\right)T^{0k}_a(r)=J^i_a-\epsilon^{ijk}\langle R^j_X\rangle P^k_a.
\end{equation}
Using our master formula~\eqref{master}, we find that the expectation values of the quark and gluon contributions to the total AM are given by
\begin{equation}
    \langle \uvec J_a\rangle=\left(\uvec{\ucal R}\times\uvec p+\frac{\uvec s}{2}\right)A_a(0) + \frac{p^0}{2m} \left( \uvec s - \frac{\uvec p(\uvec p\cdot \uvec s)}{p^0(p^0 + m)} \right)B_a(0).
\end{equation}
It follows then from Eqs.~\eqref{EVmom} and~\eqref{spinop} that
\begin{equation}
    \langle \uvec S_{X,a}\rangle=\left[(\uvec{\ucal R}-\langle\uvec R_X\rangle)\times\uvec p+\frac{\uvec s}{2}\right]A_a(0) + \frac{p^0}{2m} \left( \uvec s - \frac{\uvec p(\uvec p\cdot \uvec s)}{p^0(p^0 + m)} \right)B_a(0).
\end{equation}
Finally, using the results in Table~\ref{tab1} we obtain the key result of this work
\begin{equation}\label{key}
    \langle\uvec S_{X,a}\rangle=\langle\uvec S_X\rangle A_a(0)+\langle\uvec S_M\rangle B_a(0).
\end{equation}
Summing over the quark and gluon contributions, we recover the constraints $\sum_a A_a(0)=1$ and $\sum_aB_a(0)=0$ arising from Poincar\'e symmetry. We show in the following that various spin sum rules discussed in the literature simply correspond to particular cases of this remarkable simple-looking formula. 

\subsubsection{Helicity sum rule}

First, if we restrict ourselves to the longitudinal component, we find the unique situation that we get a sum rule \emph{independent} of the choice of position vector, i.e. a sum rule with the quark and gluon contributions given by
\begin{equation}
    \langle\uvec S_{X,a}\rangle\cdot\hat{\uvec p}=\frac{\uvec s\cdot\hat{\uvec p}}{2}\left[A_a(0)+B_a(0)\right].
\end{equation}
For a longitudinally polarized nucleon, we have in particular $\uvec s\cdot\hat{\uvec p}=1$. This is nothing else than the helicity sum rule derived in Refs.~\cite{Ji:1996ek,Ji:1997pf}. Since the choices $X=E,M,c$ all give the same result, this explains why the question of the center of the nucleon did not attract much attention in the literature.

\subsubsection{Canonical spin sum rule}

If we choose to work with the canonical position $\langle\uvec R_c\rangle=\uvec{\ucal R}$, we simply have to set $X=c$ in Eq.~\eqref{key}. The quark and gluon contributions to the canonical spin are then given by
\begin{equation} \label{Jexp}
\la \uvec S_{c,a}\ra = \frac{\uvec s}{2}\,A_a(0) + \frac{p^0}{2m} \left(\uvec s - \frac{\uvec p(\uvec p\cdot \uvec s)}{p^0(p^0 + m)}\right)B_a(0).
\end{equation}
This expression was derived for the first time in Ref.~\cite{Leader:2011cr} by setting the origin at the center of the wave packet (i.e. the relativistic center of spin), and hence computing $\langle\uvec J_a\rangle_{\uvec 0,\uvec p,\uvec s}=\langle\uvec S_{c,a}\rangle_{\uvec 0,\uvec p,\uvec s}$ from Eq.~\eqref{standardform}. Our phase-space approach generalizes in a straightforward way the derivation to an arbitrary choice of origin, i.e. to arbitrary $\uvec{\ucal R}$, thanks to Eq.~\eqref{genform}.

\subsubsection{Covariant spin sum rule}

If we are interested in the internal AM about the relativistic center of mass, it suffices to set $X=M$ in Eq.~\eqref{key}. We then find
\begin{equation}
   \langle\uvec S_{M,a}\rangle=\frac{p^0}{2m} \left( \uvec s - \frac{\uvec p(\uvec p\cdot \uvec s)}{p^0(p^0 + m)} \right)\left[A_a(0)+B_a(0)\right],
\end{equation}
This is the only case where the contributions associated with the $A_a$ and $B_a$ form factors have the same momentum dependence. In particular, even though $\uvec S_{M,a}$ is not conserved, its expectation value is proportional to $\langle\uvec S_M\rangle$ and hence transforms in a covariant way. This has to do with the fact that $\langle R^\mu_M\rangle$ transforms as a Lorentz four-vector.

Focusing on the transverse components, we get
\begin{equation}
   \langle\uvec S_{M\perp,a}\rangle=\gamma\, \frac{\uvec s_\perp}{2}\left[A_a(0)+B_a(0)\right]
\end{equation}
with $\gamma=p^0/m$. This result was derived recently in Ref.~\cite{Ji:2020hii} by computing $\langle J^i_a\rangle_{\uvec 0,\uvec p,\uvec s}=\epsilon^{ijk}\langle \int\ud^3r\, r^j T^{0k}_a(r)\rangle_{\uvec 0,\uvec p,\uvec s}$ from Eq.~\eqref{standardform} like in Ref.~\cite{Leader:2011cr}, but this time discarding terms proportional to $p^j$ or $p^k$ in $\langle \int\ud^3r\, r^j T^{0k}_a(r)\rangle_{\uvec 0,\uvec p,\uvec s}$. The authors motivated the latter prescription by the fact that the Pauli-Luba\'nski pseudovector $\langle W^\mu\rangle=\frac{1}{2}\,\epsilon^{\mu\alpha\beta\lambda}\langle M_{\alpha\beta}\rangle p_\lambda$ does not receive any contribution from terms proportional to $p^\alpha$ or $p^\beta$ in $\langle M^{\alpha\beta}\rangle$, which were accordingly interpreted as the ``contribution coming from the center-of-mass motion'' of the nucleon. Unfortunately, we did not find in Ref.~\cite{Ji:2020hii} a clear definition of the center of mass, nor a direct calculation of the contribution from its motion. Moreover, the prescription applies only at the level of expectation values and requires therefore a more rigorous justification from the operator level.

Our phase-space approach combined with the discussion about the relativistic definition of the center of the nucleon from Sec.~\ref{sec:spin} justifies and clarifies a posteriori the above ad hoc prescription. Indeed, if we consider the expectation value of spatial moments about the relativistic center of mass, we obtain from the master formula~\eqref{master}
\beq\label{momentCM}
\begin{aligned}
\langle\int\ud^3 r\left(r^i-\langle R^i_M\rangle\right)T^{\mu\nu}_a(r)\rangle&= \left(\ucal R^i-\langle R^i_M\rangle-\frac{\epsilon^{0i\alpha\beta}\ucal S_\alpha p_\beta}{2m(p^0+m)}\right)\frac{p^\mu p^\nu A_a(0)+m^2g^{\mu\nu}\bar C_a(0)}{p^0}\\
&\quad+(\delta^i_\lambda-\delta^0_\lambda\,\tfrac{p^i}{p^0})\,\frac{p^{\{\mu}\epsilon^{\nu\}\lambda\alpha\beta}\ucal S_\alpha p_\beta}{p^0m}\,\frac{A_a(0)+B_a(0)}{2}.
\end{aligned}
\eeq
According to Table~\ref{tab1}, we can write
\begin{equation}
    \ucal R^i-\langle R^i_M\rangle=\frac{(\uvec p\times\uvec s)^i}{2m(p^0+m)}=\frac{\epsilon^{0i\alpha\beta}\ucal S_\alpha p_\beta}{2m(p^0+m)}
\end{equation}
showing that the contribution from the motion of the center of the wave packet relative to the center of mass compensates exactly the part associated with the Wigner rotation. The first line in Eq.~\eqref{momentCM} then vanishes identically and we are left with
\beq\label{correct}
\langle\int\ud^3 r\left(r^i-\langle R^i_M\rangle\right)T^{\mu\nu}_a(r)\rangle= (\delta^i_\lambda-\delta^0_\lambda\,\tfrac{p^i}{p^0})\,\frac{p^{\{\mu}\epsilon^{\nu\}\lambda\alpha\beta}\ucal S_\alpha p_\beta}{p^0m}\,\frac{A_a(0)+B_a(0)}{2}.
\eeq
In particular, for the AM we have
\beq
\epsilon^{ijk}\langle\int\ud^3 r\left(r^j-\langle R^j_M\rangle\right)T^{0k}_a(r)\rangle= \epsilon^{ijk}\left[\frac{\epsilon^{kj\alpha\beta}\ucal S_\alpha p_\beta}{2m}\,\frac{A_a(0)+B_a(0)}{2}\right],
\eeq
where the expression inside the square brackets is effectively what one obtains from applying the ad hoc prescription to $\langle\int\ud^3 r\,r^j T^{0k}_a(r)\rangle_{\uvec 0,\uvec p,\uvec s}$, as was done in Ref.~\cite{Ji:2020hii}. We note however that applying naively the ad hoc prescription to $\langle\uvec K_a\rangle_{\uvec 0,\uvec p,\uvec s}$ would give a contribution from $\bar C_a(0)$, while the correct expression derived from Eq.~\eqref{correct} does not.

\subsubsection{A new spin sum rule}

For completeness, let us finally consider the internal AM about the relativistic center of energy. Setting $X=E$ in Eq.~\eqref{key}, we get
\begin{equation}
   \langle\uvec S_{E,a}\rangle=\frac{m}{2p^0} \left(\uvec s + \frac{\uvec p(\uvec p\cdot \uvec s)}{m(p^0 + m)}\right)A_a(0) + \frac{p^0}{2m} \left(\uvec s - \frac{\uvec p(\uvec p\cdot \uvec s)}{p^0(p^0 + m)}\right)B_a(0),
\end{equation}
To the best of our knowledge, this result is derived for the first time.

Since $\uvec S_E=\uvec W/P^0$, one may wonder whether this new spin sum rule is related to the sum rule based on the Pauli-Luba\'nski pseudovector proposed in Ref.~\cite{Ji:2012vj}. Beside the factor of $1/P^0$, it appears that the decomposition into quark and gluon contributions is in fact different. In Ref.~\cite{Ji:2012vj} the Pauli-Luba\'nski pseudovector is decomposed as
\begin{equation}
    W^\mu=\sum_a \mathcal W^\mu_a
\end{equation}
with the quark and gluon contributions defined as\footnote{Note that $\mathcal W^\mu_a$ is not the Pauli-Luba\'nski pseudovector of the quark subsystem since the latter would involve $P^\mu_a$ instead of $P^\mu$.}
\begin{equation}
    \mathcal W^\mu_a=\frac{1}{2}\,\epsilon^{\mu \alpha \beta \lambda} (M_a)_{\alpha\beta} P_{\lambda}.
\end{equation}
In other words, one has
\begin{equation}
    \frac{\uvec{\mathcal W}_{\!a}}{P^0}=\uvec J_a+\frac{\uvec K_a\times\uvec P}{P^0} 
\end{equation}
whereas we defined
\begin{equation}
    \uvec S_{E,a}=\uvec J_a+\frac{\langle\uvec K\rangle\times\uvec P_a}{P^0}
\end{equation}
as follows from Eqs.~\eqref{spinop} and~\eqref{id}. Clearly, the quark and gluon contributions to $\langle\uvec S_E\rangle$ are defined relative to the center of energy of the whole system and hence involve $\langle \uvec K\rangle$. In contrast, $\langle\uvec{\mathcal W}_{\!a}\rangle$ involves only $\langle \uvec K_a\rangle$ which is related to the energy dipole moment arising from the quark or gluon subsystem~\cite{Lorce:2018zpf}. This explains in particular why $\bar C_a(0)$ appears in $\langle \uvec{\mathcal W}_{\!\perp,a}\rangle$~\cite{Leader:2012ar,Hatta:2012jm,Harindranath:2013goa,Leader:2013jra} and not in $\langle \uvec S_{E\perp,a}\rangle$.

\section{Conclusions}\label{sec:conclusions}

A spin sum rule is usually understood as a decomposition of the angular momentum about the \emph{center} of the nucleon into quark and gluon contributions. While the longitudinal spin or helicity sum rule is well established, different transverse spin sum rules have been proposed in the literature. In this paper we show that there is no contradiction between the published results, since they correspond to different definitions of what constitutes the center of the nucleon. Transverse angular momentum is always more complicated in a relativistic theory because, unlike longitudinal angular momentum, it commutes neither with longitudinal boosts nor momentum. 

In the literature, spin sum rules are often obtained from the expectation value of the angular momentum operator. The latter has to be treated with care since a naive calculation using momentum eigenstates leads to ambiguous or divergent results. Some papers also make use of the Pauli-Luba\'nski pseudovector but have trouble with providing a physical meaning to the contribution coming from the boost generators, and discard terms asserted to arise from the motion of the center of mass without a clear definition of the latter.

In this work, we stress the importance of defining clearly the center of the nucleon. We review the various possibilities offered in the relativistic context (namely the relativistic centers of energy, mass and spin), and remind in passing that boost generators provide the information about the position of the center of energy. We then define expectation values from a phase-space perspective. This allows us to avoid ambiguities encountered in the standard approach, and to compute for the first time the contribution from the motion of the center of the nucleon. 

Our key result is found in Eq.~\eqref{key} which, together with Table~\ref{tab1}, gives the generic expression for the quark and gluon contributions to the nucleon spin, based on the three possible definitions of the center of the nucleon. Projecting on particular spin components with particular choices for the center of the nucleon, we reproduce several spin sum rules reported in the literature. In particular, we show that the new sum rule obtained by Ji and Yuan~\cite{Ji:2020hii} is not in contradiction with the older sum rule derived by Leader~\cite{Leader:2011cr}, but simply relies on a different definition for what is meant by internal angular momentum or spin. Namely, if one requires spin to generate rotations about the center of the nucleon, then one is forced to work with the canonical spin operator, leading to the Leader sum rule. If one requires instead spin to transform in a covariant way, then one is forced to work with the angular momentum about the relativistic center of mass, leading to the Ji and Yuan sum rule. We note that a third sum rule can be obtained based on the angular momentum about the relativistic center of energy or inertia, which has the simplest expression in terms of the Poincar\'e generators.

\begin{acknowledgements}
I am very grateful to Elliot Leader for many valuable comments and suggestions about the manuscript.
\end{acknowledgements}

\color{black}


\end{document}